\documentclass{emulateapj}

\newcommand{\beq}{\begin{equation}}
\newcommand{\beqa}{\begin{eqnarray}}
\newcommand{\eeq}{\end{equation}}
\newcommand{\eeqa}{\end{eqnarray}}

\newcommand{\bfx}{\mathbf{x}} 
\newcommand{\bfk}{\mathbf{k}}

\shorttitle{non-Gaussian error in likelihood analysis}
\shortauthors{Takahashi et al.}

\begin{document}

\title{Non-Gaussian Error Contribution to
Likelihood Analysis of the Matter Power Spectrum}

\author{Ryuichi Takahashi\altaffilmark{1},
Naoki Yoshida\altaffilmark{3},
Masahiro Takada\altaffilmark{3},
Takahiko Matsubara\altaffilmark{2}, 
Naoshi Sugiyama\altaffilmark{2,3}, 
Issha Kayo\altaffilmark{3}, 
Takahiro Nishimichi\altaffilmark{4},
Shun Saito\altaffilmark{4},
Atsushi Taruya\altaffilmark{3,5}}

\affil{\altaffilmark{1} Faculty of Science and Technology, Hirosaki
 University, 3 bunkyo-cho, Hirosaki, Aomori, 036-8561, Japan}
\affil{\altaffilmark{2} Department of Physics,
 Nagoya University, Chikusa, Nagoya 464-8602, Japan}
\affil{\altaffilmark{3} Institute for the Physics and Mathematics of the Universe,
The University of Tokyo, 5-1-5 Kashiwa-no-ha, Kashiwa, Chiba 277-8568, Japan}
\affil{\altaffilmark{4} Department of Physics, School of Science,
 The University of Tokyo, Tokyo 113-0033, Japan}
\affil{\altaffilmark{5} Research Center for the Early Universe,
 The University of Tokyo, Tokyo 133-0033, Japan}

\begin{abstract}

 We study the sample variance of the matter power spectrum 
for the standard $\Lambda$ Cold Dark Matter universe.
We use a total of 5000 cosmological $N$-body simulations
to study in detail the distribution of best-fit cosmological
parameters and the baryon acoustic peak positions.
The obtained distribution is compared with the results 
from the Fisher matrix analysis with and without 
including non-Gaussian errors. For the Fisher matrix analysis, we 
compute the derivatives
of the matter power spectrum with respect to cosmological
parameters using directly full nonlinear simulations.
We show that the non-Gaussian errors increase the unmarginalized 
errors by up to a factor $5$ for $k_{\rm max}=0.4h/$Mpc 
if there is only one free parameter provided other parameters 
are well determined by external information. 
On the other hand, for multi-parameter fitting, the impact of the
non-Gaussian errors is significantly mitigated due to severe parameter 
degeneracies in the power spectrum. 
The distribution of
 the acoustic peak positions is well described by a Gaussian
 distribution, with its width being consistent with the statistical
interval predicted from the Fisher matrix. 
We also examine systematic bias in the best-fit parameter 
due to the non-Gaussian errors.
The bias is found to be smaller than the $1 \sigma$ statistical error 
for both the cosmological
 parameters and the acoustic scale positions.
\end{abstract}


\section{Introduction}

The baryon acoustic oscillation (BAO) is imprinted in the distribution
of galaxies as is found in the temperature fluctuations in the cosmic
 microwave background. 
The acoustic length scale is determined by the sound horizon
 of the photon-baryon fluid at recombination epoch; it can thus be used as a
 standard ruler which provides us with a robust method to measure 
distance scales out to essentially any epoch
 (e.g., Eisenstein et al. 1998; Blake \& Glazebrook 2003; 
 Seo \& Eisenstein 2003; Matsubara 2004; Guzik et al. 2007). 
Using the observed distance-redshift relation, we can obtain an accurate cosmic expansion
history, which in turn gives strong constraints on the
 nature of dark energy.
The large-area galaxy surveys such as two-degree Field Survey (2dF) and Sloan
 Digital Sky Survey (SDSS) detected the BAO signature in the galaxy
 distribution (Cole et al. 2005; Eisenstein et al 2005;
 Percival et al. 2007; Okumura et al. 2008;
 Gaztanaga et al. 2008; Sanchez et al. 2009).
The latest result of the SDSS DR7 showed
a constraint on the distance to a redshift $z=0.28$ 
 within $2.7 \%$ accuracy
 (Percival et al. 2009; Reid et al. 2009; Kazin et al. 2009).
Future and ongoing surveys such as the Baryon Oscillation 
 Spectroscopic Survey (BOSS)\footnote{http://www.sdss3.org/cosmology.php},
 the Hobby-Eberly Dark Energy Experiment
 (HETDEX)\footnote{http://hetdex.org/}, and
 the WiggleZ surveys\footnote{http://wigglez.swin.edu.au/Welcome.html}
 will measure the distance to higher redshifts within a few percent accuracy.

The BAO signature appears as a small wiggle pattern in the galaxy 
 power spectrum.
Since the amplitude of BAO wiggle is very small ($\sim$ a few
 percent), rather accurate theoretical models are needed.
Especially, in order to determine the distance within a percent accuracy
 for the planned or ongoing surveys,
 we need to be able to predict the acoustic scale with much higher
 accuracies ($\sim 0.1 \%$).
However, there are complicated astrophysical processes such as the non-linear
 gravitational evolution, scale-dependent bias of galaxies, redshift space
 distortion, and the effect of massive neutrino.
Many authors tackled these problems using numerical simulations
 (Meiksin et al. 1999; Seo \& Eisenstein 2005; Huff et al. 2007;
  Smith, Scoccimarro \& Sheth 2007, 2008; Angulo et al. 2008; 
  Takahashi et al. 2008; Seo et al. 2008; Nishimichi et al. 2009; 
  Kim et al. 2009; Heitmann et al. 2009)
and analytical perturbation theories   
 (Crocce \& Scoccimarro 2006, 2008; Jeong \& Komatsu 2006, 2009;
  Nishimichi et al. 2007; McDonald 2007; Matarrese \& Pietroni 2007, 2008;
  Eisenstein et al. 2007;
  Pietroni 2008; Matsubara 2008a,b; Taruya \& Hiramatsu 2008;
  Takahashi 2008; Nomura et al. 2008; Rassat et al. 2008;
  Sanchez et al. 2008; Padmanabhan \& White 2008;
  Saito, Takada \& Taruya 2009; Shoji, Jeong \& Komatsu 2009; 
  Taruya et al. 2009; Montesano et al. 2010).

 It is crucial to use not only accurate power spectra but also accurate
 covariance matrices in order to determine cosmological parameters 
from the galaxy power spectrum.
If the matter density fluctuations obey a Gaussian distribution,
 the covariance matrix has
 only diagonal element and the relative error is simply given by the
 square root of the number of modes in the survey area
 (e.g. Feldman, Kaiser \& Peacock 1994).
However, when the density fluctuations grow to the non-linear regime,
 the mode coupling between different wavenumbers 
 generates non-zero off-diagonal elements, and the so-called non-Gaussian error
 is induced (e.g. Scoccimarro, Zaldarriaga \& Hui 1999; Meiksin \& White 1999).
Rimes \& Hamilton (2005,2006) first pointed out that there is little
 information contained in the power spectrum at quasi-nonlinear regime
 ($k=0.2-0.8h/$Mpc) due to the non-Gaussian error
 (see also Hamilton, Rimes \& Scoccimarro 2006;
 Neyrinck, Szapudi \& Rimes 2006; Neyrinck \& Szapudi 2007,2008;
 Lee \& Pen 2008; Neyrinck, Szapudi \& Szalay 2009; Lu, Pen \& Dore 2009;
 Sato et al. 2009).
For weak lensing (cosmic shear) analysis, the non-Gaussian error 
contribute the total error substantially 
 (Cooray \& Hu 2001; Sefusatti et al. 2006; Dore, Lu \& Pen 2009; 
 Takada \& Jain 2009; Pielorz et al. 2009),
 but also may systematically shift the best fitting parameter
 (Hartlap et al. 2009; Ichiki et al. 2009). 
Especially, if there are a small number of parameters to be determined, 
the non-Gaussianity affects the errors significantly
 (see the discussion in Takada \& Jain 2009). 

In our previous paper (Takahashi et al. 2009, hereafter T09),
 we used $5000$ cosmological simulations to obtain
 the accurate covariance matrix of the matter power spectrum.
This is a largest number of realizations ever done for the
 cosmological N-body simulation.
We studied the non-Gaussian error contribution to  
 the signal-to-noise ratio for the measurement of the power spectrum,
 and found that the non-Gaussian error is important at small length-scale
 $k>0.2h/$Mpc. 
In this paper, we further investigate the non-Gaussian error contribution
 to the cosmological parameter estimation and the 
best-fit values
 using the $\chi^2$ likelihood analysis. 
We calculate the distribution of 
the best-fit parameters among the $5000$ realizations, and compare
 it with the results using the Fisher matrix analysis. 
We also study the distribution of the acoustic scale positions
 among the realizations.
Our results in this paper can be used not only for the BAO
 analysis but also for the more general issue in the likelihood
 analysis of the non-linear matter power spectrum.

Throughout the present paper, we adopt the standard $\Lambda$CDM model
with matter density $\Omega_{m} =0.238$, baryon density
$\Omega_{\rm b}=0.041$, dark energy density $\Omega_{\rm w}=0.762$
 with equation of state $w=-1$,
spectral index $n_{\rm s}=0.958$, amplitude of fluctuations $\sigma_8=0.76$, 
and expansion rate at the present time $H_{0}=73.2$km s$^{-1}$ Mpc$^{-1}$,
consistent with the 3-year WMAP results (Spergel et al. 2007).

\section{Matter Power Spectrum and Its Covariance Matrix from Numerical 
 Simulations}

We follow the gravitational evolution of $256^3$ collisionless dark matter
 particles in a volume of $1000 h^{-1} {\rm Mpc}$ on a side using the
 cosmological simulation code Gadget-2
 (Springel, Yoshida \& White 2001; Springel 2005).
We generate initial conditions following the standard Zel'dovich
 approximation using the matter transfer function calculated by CAMB
 (Code for Anisotropies in the Microwave Background: Lewis, Challinor
 \& Lasenby 2000; also see Seljak \& Zaldarriaga 1996).
The initial redshift is set to be $z=20$. We use outputs at $z=3,1$ and $0$.
To calculate the density fluctuations,
 we assign the N-body particles onto a $512^3$ rectangular grid using the
 cloud-in-cell scheme.
Then we perform the Fourier transform and
 calculate the power spectrum in both real space and redshift space.
We run $5000$ Particle-Mesh(PM) simulations to follow the
 non-linear evolution of the power spectrum and its covariance matrix
 in detail. 
We have checked that the power spectra of our simulations agree well
 with the result of the higher resolution TreePM simulation, within
 $1 (3) \%$ for $k < 0.2(0.4)h/$Mpc (here the Nyquist wavenumber is
 $k=0.8h/{\rm Mpc}$).
If the initial redshift is set to be higher, $z=50$, the results agree 
 within $2(10) \%$ for $k < 0.2(0.4)h/$Mpc\footnote{The agreement is
 achieved in real space. In redshift space, PM simulations somewhat 
 underestimate the power spectrum at small scales
 ($k=0.4h/{\rm Mpc}$) by $20 \%$.}. 
This is a sufficient accuracy 
for our purpose, which is to investigate the
 non-linear evolution of the power spectrum at BAO scales.

Denoting $\hat{P}_i(k)$ as the power spectrum computed from 
the $i$-th realization,
the ensemble averaged power spectrum is estimated from the mean of the
power spectra between 5000 realizations: 
\begin{equation}
  \bar{P}(k) = \frac{1}{N_{\rm r}} \sum_{i=1}^{N_{\rm r}}
   \hat{P}_i(k), 
\end{equation}
where $N_r=5000$, the number of our realizations. Similarly, the
covariance matrix between the spectra of
 $k_1$ and $k_2$ 
is estimated as
\beq
  {\rm cov}(k_1,k_2)= \frac{1}{N_{\rm r}-1} \sum_{i=1}^{N_{\rm r}}
    \left[ \hat{P}_i(k_1) - \bar{P}(k_1) \right]
    \left[ \hat{P}_i(k_2) - \bar{P}(k_2) \right].  \nonumber \\
\eeq
%
The accuracy of the covariance is analytically estimated for
the Gaussian density fluctuations (see Appendix). For example, the
relative errors in the diagonal covariance terms are found to scale with
the number of realizations as $(2/N_r)^{1/2}$; our 5000 simulations
provide a few percent accuracy.
Clearly, our study achieves an unprecedented accuracy of the covariance 
matrix estimation on BAO scales. 

The power spectrum covariance 
is formally expressed as a sum of the two contributions, the Gaussian and
non-Gaussian terms
(e.g. Scoccimarro, Zaldarriaga \& Hui 1999; Meiksin \& White 1999):
\beqa
  {\rm cov}(k_1,k_2) \equiv
  \left< \left( \hat{P}(k_1) - P(k_1) \right) \left( \hat{P}(k_2) - P(k_2)
 \right) \right>  \nonumber \\
 =  \frac{2}{N_{k_1}} P^2(k_1) \delta^K_{k_1,k_2} + \frac{1}{V}
 \int_{|\bfk{}'_1|\in k_1} \int_{|\bfk{}'_2|\in k_2}
 \!\!\frac{d^3 \bfk_1'}{V_{k_1}} \frac{d^3 \bfk_2'}{V_{k_2}} \nonumber \\
 \times ~T(\bfk_1',-\bfk_1',\bfk_2',-\bfk_2').
\label{cov}
\eeqa
The first term arises from the Gaussian fluctuations, while the second term
 is the non-Gaussian error arising from the mode coupling during
 the non-linear evolution. 
Here, $P(k)=\langle \hat{P}(k) \rangle$ is the mean power spectrum, 
 $T$ is the trispectrum, the integral is done over the shell of the 
 radius $k_{1,2}$ with the width $\Delta k$ in Fourier space,
 and $V_{k_{1,2}}$ is the volume of the shell
 given by $V_{k} = 4 \pi k^2 \Delta k$.
The expression in Eq.(\ref{cov}) depends on the bin width, 
 the first term is proportional to $1/(V \Delta k)$, while the second term
 is $\propto 1/V$.
Hence, for the 
finer bin width, the impact of the Gaussian term becomes relatively enhanced. 
Note however that 
the parameter estimation shown in
the following 
is independent of the bin width.
Throughout this paper, 
the bin width is set to $\Delta k = 0.01h/$Mpc.\footnote{We also try the
 half bin-width, $\Delta k = 0.005h/$Mpc, but our results in the
 following sections are almost same.}

In this paper, we do not consider another non-Gaussian term in Eq.(\ref{cov})
 arising from the finite survey volume (the so-called beat-coupling
 effect; Rimes \& Hamilton 2006).
Fluctuations with wavelength larger than the survey region may 
contribute to the covariance on smaller scale. 
Although this effect can increase the covariance by over ten percent (T09), 
the main conclusions we draw in the present paper remain robust
to the uncertainty.


\section{Effects of Non-Gaussian Error on Likelihood Analysis
 of Cosmological Parameters}


\subsection{Parameter Estimation for Cosmological Parameters}


In this section, we study the effects of the non-Gaussian errors on
 the cosmological parameter estimation 
given the power spectrum measured from a hypothetical survey of
(1$h^{-1}$Gpc)$^3$ volume. 
We use the Fisher 
matrix formalism
to estimate the accuracy of parameter estimation\footnote{Here,
 we do not consider the Alcock-Paczynski effect (Alcock \& Paczynski 1979)
 which would affect the measurement accuracy for the dark energy.}
 (Tegmark, Taylor \& Heavens 1997):
\beq
 F_{\rm ij} = \sum_{k_{1,2}<k_{\rm max}} {\rm cov}^{-1} (k_1,k_2)
 \left. \frac{\partial P(k_1;\bfx)}{\partial \ln x_{\rm i}}
 \right|_{\rm fid} 
 \left. \frac{\partial P(k_2;\bfx)}{\partial \ln x_{\rm j}} 
 \right|_{\rm fid}.
\label{fisher}
\eeq
where $x_{\rm i}$ denotes cosmological parameters, and the partial
derivative such as $\partial P/\partial x_{\rm i}$ is evaluated around
the fiducial model. We include 5 parameters (therefore ${\rm
i}=1,2,\dots,5$): the primordial power spectrum 
parameters,\footnote{We assume the primordial power spectrum given as 
$P_0(k) \propto A_s^2 (k/k_0)^{n_s}$, 
where the pivot wavenumber 
 $k_0$ is set to $k_0=0.002/$Mpc as employed by 
Komatsu et al. (2009).}
the normalization parameter $A_s$ (not $\sigma_8$) and 
 the spectral index $n_{\rm s}$, 
the baryon density $\Omega_{\rm b} h^2$,
 the dark matter density $\Omega_{\rm c} h^2$, and the 
dark energy equation of state parameter $w$. 
We assume a flat universe throughout the present paper. 
In Eq.~(\ref{fisher})
we use $\ln x_{\rm i}$ (not $x_{\rm i}$) as the variables such that 
the Fisher matrix gives the relative accuracy of a given parameter
estimation: the marginalized error is then given as
 $\Delta x_{\rm i}/x_{\rm i}=(F^{-1})^{1/2}_{\rm ii}$.
Note that for $w$, which has a negative value for the fiducial value, we
simply compute $w\partial \ln P/\partial w $ for the derivative.  
From Eqs.(\ref{cov}) and (\ref{fisher}), the estimation error
 $\Delta x_{\rm i}/x_{\rm i}$ is inversely proportional to the survey
 volume as
$\Delta x_{\rm i}/x_{\rm i} \propto V^{-1/2}$.

\begin{figure*}
\epsscale{0.8}
\plotone{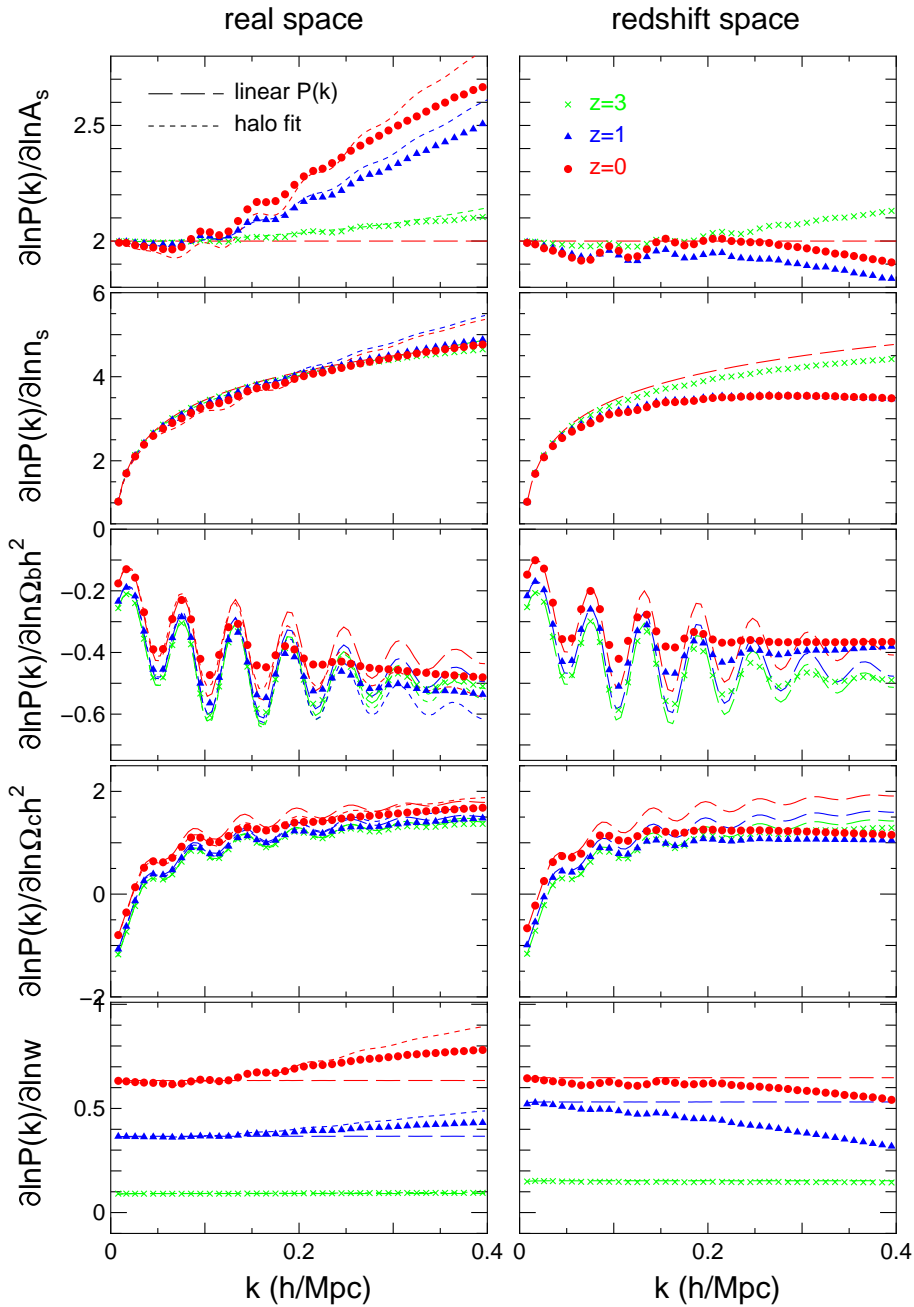}
\caption{The derivatives of non-linear power spectrum with respect to
 cosmological parameters $(A_s,n_s,\Omega_b h^2,\Omega_c h^2,w)$ in 
real-space (left panel) and redshift-space (right).
The cross, triangle and circle symbols show the simulation results at 
redshifts $z=3$, 1 and 0, respectively. 
The dashed curves are the linear theory predictions, 
 the dotted curves are for the halofit.
}
\label{fig_deriv_pk}
\vspace*{0.5cm}
\end{figure*}

We need to compute the derivatives of the power spectrum
to compute the Fisher matrix in Eq.~(\ref{fisher}). 
For each cosmological parameter, we ran simulations with
one parameter slightly varied, while fixing other parameters
to the fiducial values. We then compute the derivatives by the two-side
 differences of steps $\Delta x_i/x_i=\pm 0.05$. We use 40 realizations
 are used for each parameter variation. 
Fig.\ref{fig_deriv_pk} shows the derivatives of the power spectrum with
 respect to each cosmological parameter in real space (left column) and 
 in redshift space (right column), respectively.
The symbols 
are our simulation results: 
the red circles are for $z=0$, the blue triangles for $z=1$, and
the green crosses for $z=3$, respectively. The sensitivity of the power
spectrum to cosmological parameters appears differently between
in real- and redshift-space, due to the redshift distortion effects. 
For example, for a model with higher power spectrum normalization
i.e. larger $A_s$, the power spectrum amplitudes are increasingly
enhanced at larger $k$ due to the stronger nonlinearities in real space. 
In redshift space, however, the enhancement is significantly suppressed
by the stronger finger-of-God effect, which arises from random velocity
dispersion of dark matter particles in nonlinear objects. 
The relative amplitude of baryon acoustic oscillations is 
enhanced with increasing the baryon density.
Finally, a change in $w$ affects the power spectrum via the effect on
the growth rate. We naively expect that the dependence 
of $P(k)$ on $w$ becomes
scale-dependent in the nonlinear regime. However, the induced 
scale-dependence is
weak and the dark energy parameter is very likely degenerated with the
galaxy bias in the measured galaxy power spectrum.

The simulation results are compared with the analytical
predictions computed from the linear theory (the dashed curve) and the
halofit (the short-dashed curve: Smith et al. 2003),
 respectively.\footnote{We also compare the Lagrangian perturbation theory
 (LPT: Matsubara 2008a) with the simulation results in the manuscript in
 previous version (Takahashi et al. arXiv:0912.1381v1).
 For the redshift-space spectrum, the LPT agrees well with the simulations
 at small $k$, but deviates significantly in the weakly nonlinear regime
 due to a too significant exponential damping, $P_{\rm LPT}(k) \propto
 \exp[-{\rm const.} \times k^2]$ (Matsubara 2008a).}
The redshift-space power spectrum derivatives are compared with the linear
theory.
In the linear power spectrum in redshift space, we use the Kaiser
 approximation and do not include the Finger-of-God term. 
All the analytical predictions agree well 
with the simulation results
at small $k$. In particular, the halofit agrees with the 
simulations to within $10\%$ accuracy. 

In our previous paper (T09), we studied the impact
 of the non-Gaussian error on
 the signal-to-noise ratio of power spectrum measurement: 
\beq
 \left( \frac{S}{N} \right)^2 = \sum_{k_{1,2}<k_{\rm max}}
 {\rm cov}^{-1} (k_1,k_2) P(k_1) P(k_2).
\label{snr}
\eeq
It was found that, in the linear regime, 
the $S/N$ keeps increasing with increasing the maximum wavenumber $k_{\rm
max}$ as 
$(S/N) \propto k_{\rm max}^{3/2}$. 
However the $S/N$ saturates at some $k_{\rm max}$ in the weakly
nonlinear regime, and stays nearly constant at larger
$k_{\rm max}>0.2h/{\rm{Mpc}}$ due to the non-Gaussian errors.
From these results one may naively guess that the parameter estimation is
also significantly affected by the non-Gaussian errors when the power
spectrum information to the larger $k_{\rm max}$ is included. In the
following, we will study the impact of the non-Gaussian errors on the
parameter estimation. 

\begin{figure*}
\epsscale{0.8}
\plotone{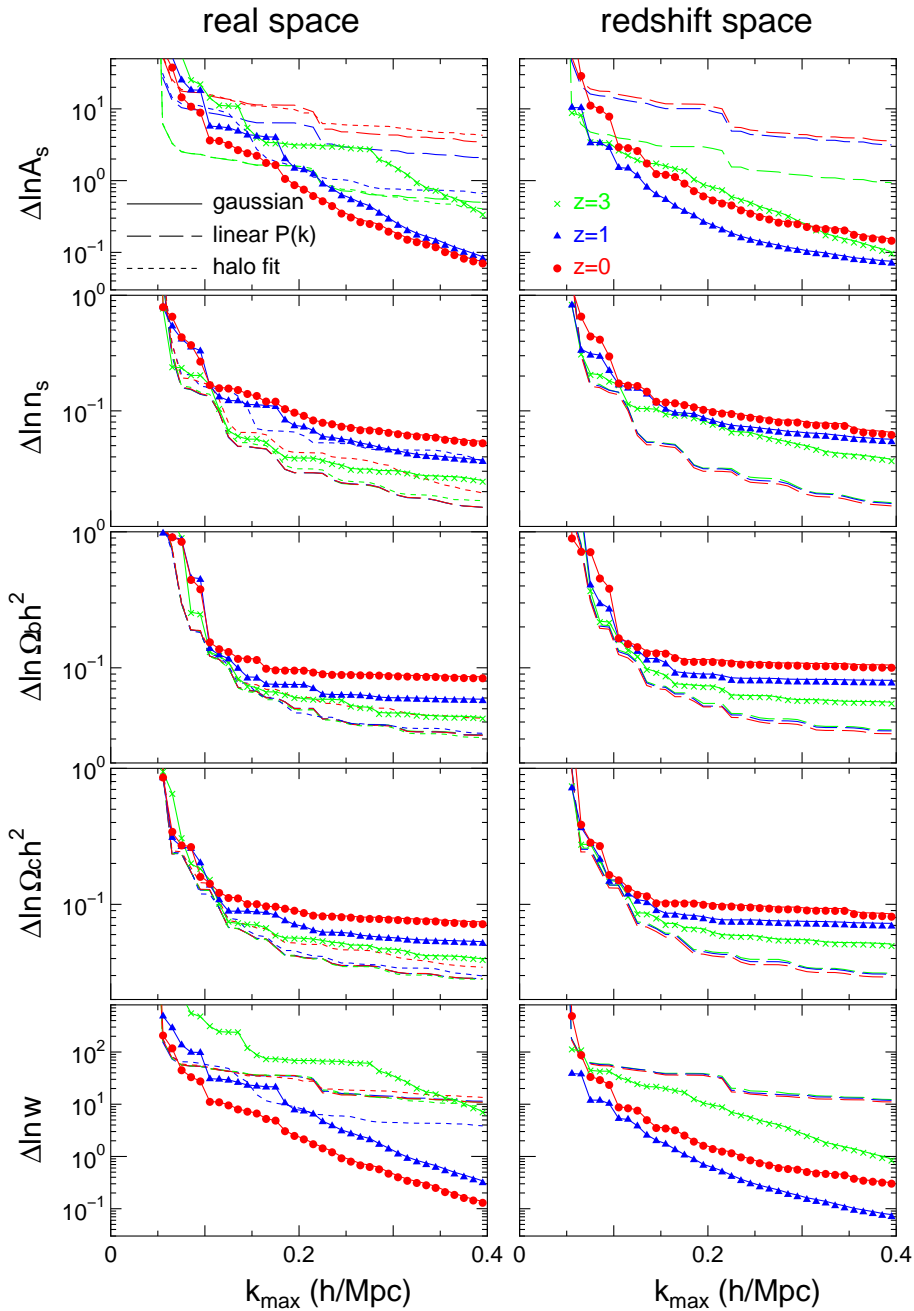}
\caption{The marginalized errors of each parameter when including the 
power spectrum information up to the maximum wavenumber $k_{\rm max}$ in
 the horizontal axis. The left- and right panels show the results for the
 real- and redshift-space, respectively. The symbols in each panel are
 as for the previous figures. 
The solid curves which lie almost on top of the symbols show the results
obtained only by including the Gaussian errors in the Fisher analysis. 
The agreement indicates that 
the Gaussian error assumption is a good approximation for parameter
 estimations even at small scales (see text for the details).
The dashed and dotted curves are the analytical predictions that are
 derived using the linear theory and halofit for the power spectrum and
 the Gaussian covariance, respectively. 
}
\label{fig_para_v3}
\vspace*{0.5cm}
\end{figure*}

Fig.\ref{fig_para_v3} shows the marginalized error on each parameter, 
$\Delta x_{\rm i}/x_{\rm i}=(F^{-1})^{1/2}_{\rm ii}$,
as a function of $k_{\rm max}$, where the power spectrum information up
to a given $k_{\rm max}$ is included. 
In each panel the symbols show the simulation results including the full
covariance matrix. 
The simulation results are almost indistinguishable from the solid curves
that are computed only by including the Gaussian error covariances,
computed from simulations, in Eq.~(\ref{cov}). 
The agreement indicates that the Gaussian error assumption
actually provides a good approximation for the parameter estimation 
over scales of
interest, even though the non-Gaussian errors have a significant impact
on the $S/N$ at $k_{\rm max}>0.2h/$Mpc. 
A more quantitative interpretation of these results will be
given later. For comparison, the dashed and dotted curves show the
results obtained by using the linear theory and halofit to estimate the
power spectrum as well as the
Gaussian error covariances. These analytical predictions are far from
the simulation results due to their inaccuracies in comparison with the
simulations.
Although the power spectrum and its derivatives in the halo fit agree
 within $\sim 10 \%$ with the simulations (see Fig.\ref{fig_deriv_pk}),
 the parameter estimations are largely different as shown in
 Fig.\ref{fig_para_v3}.
This is because when calculating the inverse matrix of the Fisher matrix,
 even a small errors in the Fisher matrix generates large errors in the
 inverse matrix.

\begin{figure*}
\epsscale{0.8}
\plotone{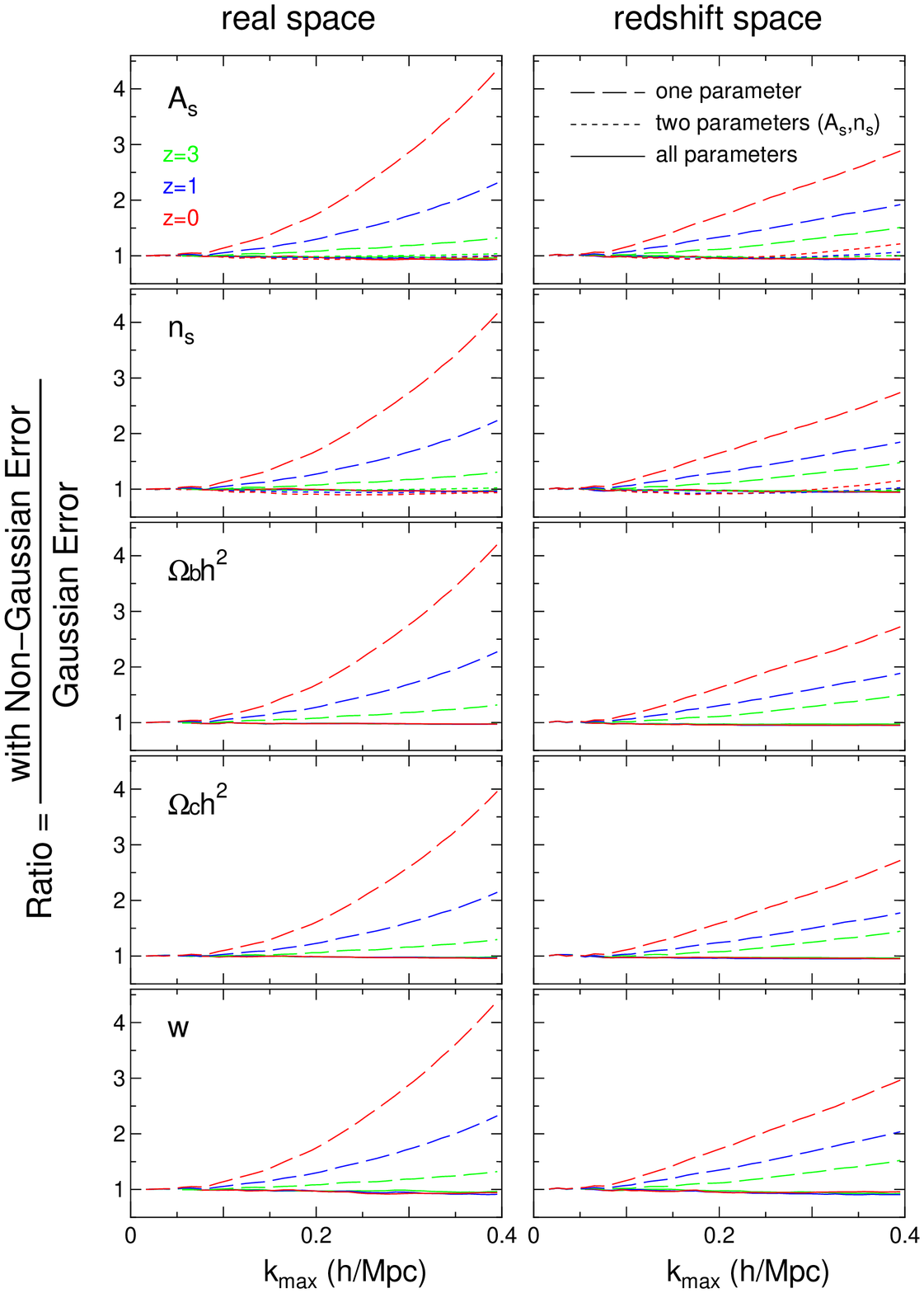}
\caption{
The ratio of the marginalized errors with and without the
 non-Gaussian errors, as a function of $k_{\rm max}$. 
The solid curves show the results for our fiducial set of five cosmological
parameters as shown in each panel. For comparison the
 dashed curves show the results for the unmarginalized errors or
 equivalently for the case of one parameter fitting in each panel. Also
in the panels for $A_s$ and $n_s$ the dotted curves show the
 results for the two parameter fitting of ($A_s$, $n_s$), which appear 
 to be between the solid and dashed curves. 
It is clear that the non-Gaussian errors degrade the unmarginalized
 errors by up to a factor 5. 
}
\label{fig_para_v4}
\vspace*{0.5cm}
\end{figure*}

Fig.\ref{fig_para_v4} shows the relative accuracies of each parameter
estimation as a function of $k_{\rm max}$. There, we
compare the results derived from the covariances with and
without the non-Gaussian error contributions. The solid curves are the
results where all the five parameters are included in the Fisher
analysis, while the dashed curve shows the unmarginalized error on each
parameter, i.e., the error is obtained by considering only one free parameter,
$\Delta x_{\rm i}/x_{\rm i} = F_{\rm ii}^{-1/2}$. In other words, the dashed curves
correspond to the case where other parameters are well constrained by
external data sets. The difference between the solid and dashed curves
is caused by the parameter degeneracies; the marginalized error
becomes same as the unmarginalized error when the parameters are
independent in the measured power spectrum. It is clear that, for the
unmarginalized errors, including the non-Gaussian covariances
degrades the parameter errors by a factor 4-5 for the redshift $z=0$, and
by a factor 2-3 for $z=1$, respectively. The level of the degradation
is similar to that of the $S/N$ as found in T09. Therefore, the impact of
non-Gaussian covariance errors is significantly mitigated by the
parameter degeneracies  
(see also, Neyrinck \& Szapudi 2007; Takada \& Jain 2009).
The degradation of $S/N$ increases the full Fisher
ellipsoid volume, and then individual parameters are not
tightly constrained due to the parameter degeneracies in 
such a high-dimension parameter space. 
As shown in Fig.\ref{fig_para_v4}, the non-Gaussian effect becomes
 negligible for the multi-parameter fitting. We comment that this
 conclusion is independent of the volume of the survey.

In the upper two panels for $A_s$ and $n_s$ the short 
dashed curves show the results for
the two parameter fitting case  $(A_s, n_s)$, which are very similar
to the solid curves. 
In reality, parameters that describe galaxy bias need to be further included.
We thus conclude that the impact of the non-Gaussian errors is
less important than the parameter degeneracies, and that 
the Gaussian covariances can
provide a good approximation to obtain the statistical uncertainty of
given parameters.

\subsection{Distribution of Best-Fit Parameters}

Nonlinear structure formation causes non-Gaussian distributions of
the power spectrum estimators at small scales, as studied in, e.g., T09, 
in detail. Here, utilizing our 5000 realizations, we quantify the
distribution of parameter estimation taking into account the
non-Gaussian covariances and the marginalization over other parameters.
To this end, we simply use the $\chi^2$-fitting analysis given as
\beqa
  \chi^2_i(\bfx) = \sum_{k_{1,2}<k_{\rm max}}
 {\rm cov}^{-1} (k_1,k_2) \left[ P(k_1;\bfx) - \hat{P}_i(k_1) \right]
 \nonumber \\
 \times \left[ P(k_2;\bfx) - \hat{P}_i(k_2) \right],
\label{chi2}
\eeqa
where $\bfx=(A_s,n_s,\Omega_b h^2, \Omega_c h^2,w)$, 
$\hat{P}_i(k)$ is the power spectrum estimator of th $i$-th realization
 and $P(k;\bfx)$ is its mean. For this analysis we simply use the
  real-space power spectrum. 

The variation in the power spectrum around the fiducial model can be
expressed as
\beq
  P(k;\bfx) \simeq P(k;\bfx_{\rm fid}) + \left. \frac{\partial P(k;\bfx)}
 {\partial \bfx} \right|_{\rm fid} \cdot \left( \bfx - \bfx_{\rm fid} \right).
\label{taylor}
\eeq  
%
Recall that the best-fit parameters are estimated by minimizing the
$\chi^2$. The best-fit parameters for the $i$-th realization
can be estimated by 
inserting Eq.~(\ref{taylor}) into Eq.~(\ref{chi2})
 (e.g. Huterer \& Takada 2005; Joachimi \& Schneider 2009):
\beqa
 \left( \bfx_{\rm bf} - \bfx_{\rm fid} \right)_i =
 \sum_j \left( F^{-1} \right)_{ij}
 \sum_{k_1,k_2<k_{\rm max}} {\rm cov}^{-1} (k_1,k_2) \nonumber \\
 \times \left[
 P_i(k_1) - P(k_1,\bfx_{\rm fid}) \right] \left.
 \frac{\partial P(k_2;\bfx)}{\partial x_j} \right|_{\rm fid}.
\label{deviation}
\eeqa
Thus the best-fit parameters are generally different from the fiducial
values depending on the distribution of $P_i$ or how $P_i$ deviates from
the ensemble average expectation $P$ at each wavenumber.
Strictly speaking, we need to vary the covariance as a function of
cosmological models in Eq.~(\ref{deviation}), but we here simply employ
the covariance for the fiducial model assuming that the variations in
the covariance are small (see Eifler, Schneider \& Hartlap 2009 for this
issue).

\begin{figure*}
\epsscale{0.8}
\plottwo{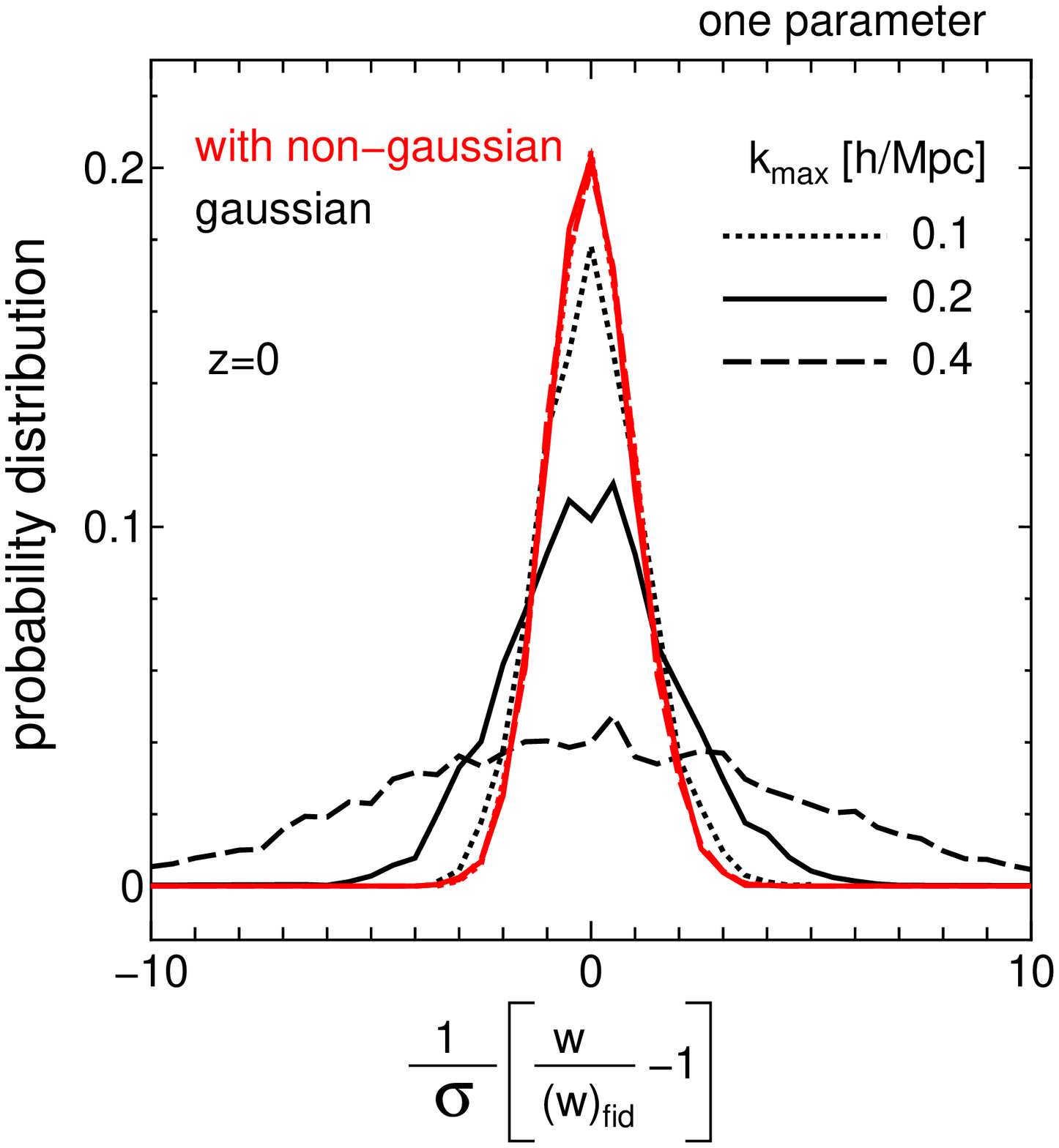}{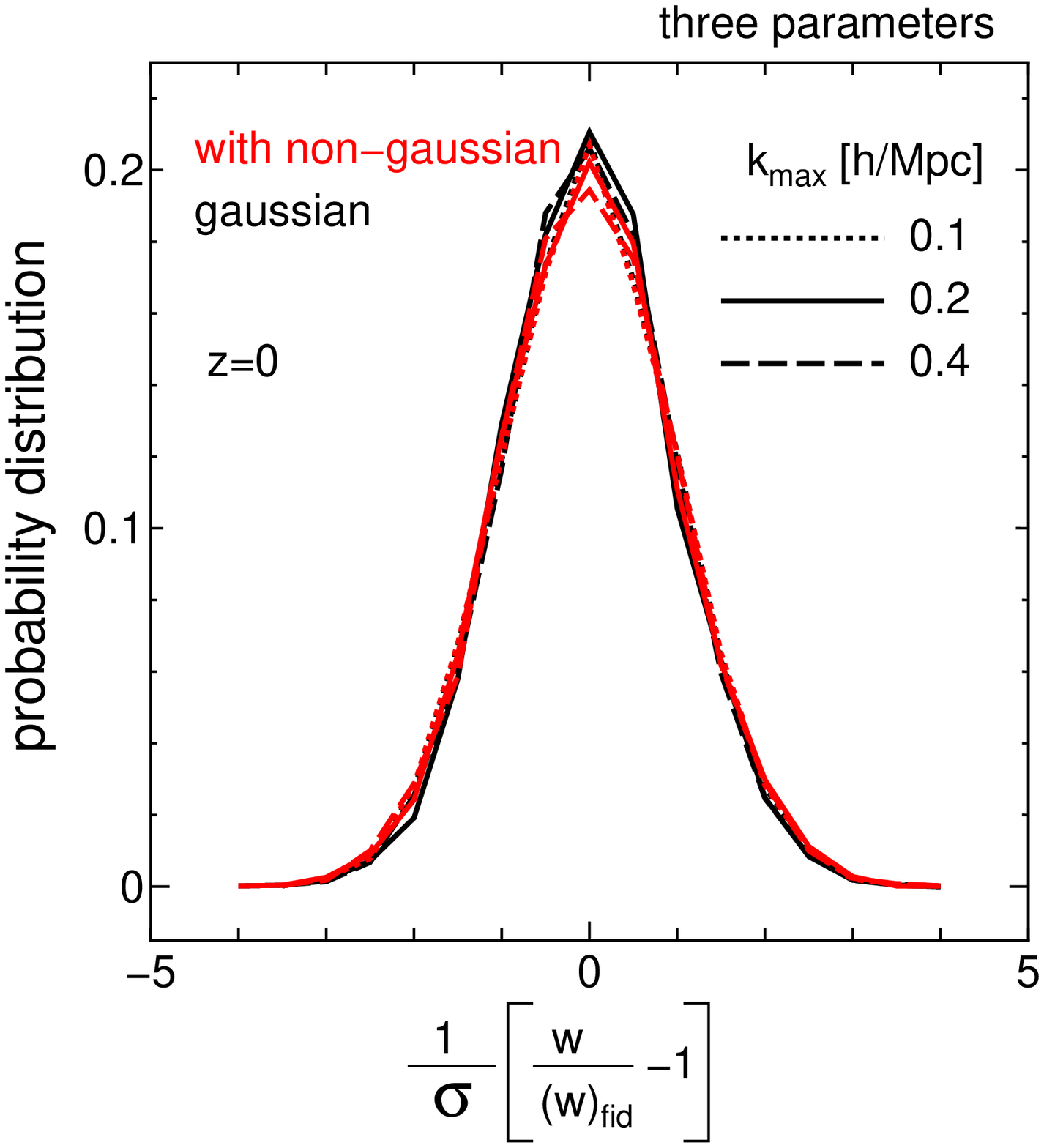}
\caption{The distribution of the best-fit parameter $w$ estimated for
 the power spectra of 
 $5000$ realizations in real space.
The red (black) curve shows the result obtained with (without)
the non-Gaussian error covariance in Eq.~(\ref{deviation}). 
The dotted, solid and dashed curves are for  
 for $k_{\rm max}=0.1$, 0.2 and $0.4h/$Mpc, respectively. 
The left panel shows the best-fit values of $w$ for one parameter
 fitting ($w$ alone), while the right panel shows the results for
 three parameter fitting ($w,A_s, n_s$). 
The horizontal axis is defined as $(w/w_{\rm fid}-1)/\sigma$, where
 $\sigma$ denotes the 1$\sigma$ confidence regions computed from the
 Fisher matrix with and without the non-Gaussian errors for the red and
 black curves, respectively. 
}
\label{dist_w}
\vspace*{0.5cm}
\end{figure*}
Fig.\ref{dist_w} shows how the best-fit values of $w$ are distributed
among 5000 realizations. Note that, as can be seen in
Eq.~(\ref{deviation}), the dark energy constraint includes the power
spectrum amplitude information in addition to the BAO features. 
The left-panel shows the result for one parameter fitting $(w)$, while
the right panel for the three parameter fitting
($w,A_s,n_s$), where the best-fit $w$ is derived by including
marginalization over  
the two parameters ($A_s,n_s$).
The latter corresponds to the case that 
the other parameters ($\Omega_c h^2,\Omega_b h^2$) are well constrained
 by external information such as 
the CMB and/or the Big Bang Nucleosynthesis (BBN). The red (black)
curve 
shows the result obtained when the non-Gaussian errors in the
Fisher matrix and the covariance in Eq.~(\ref{deviation}) are included
(not included). The corresponding best-fit parameter
deviations are plotted in the unit, $[(w/w_{\rm fid})-1]/\sigma$, where 
$\sigma$ is set to the Fisher errors with and without the non-Gaussian errors 
for the red and black curves, respectively. For example, the parameter
deviations $|(w/w_{\rm fid})-1|/\sigma\le 3 $ mean that the parameter
deviations are within 
$\pm 3\sigma$ confidence level regions. 

The distribution of the best-fit $w$ looks nearly
symmetric: the nonlinear power spectrum does not shift 
the $w$-parameter to either of negative and positive 
sides
from the fiducial value. The left panel (one-parameter fitting
case) shows that including only the Gaussian errors makes the
distribution of the  
best-fit $w$ broader than the statistical confidence region. 
Clearly, a strong evidence on $w\ne -1$ may be incorrectly derived with
high chances under the Gaussian assumptions. 
However, the red curves demonstrate that such
apparent deviations can be corrected if we properly take into account
the non-Gaussian errors for the statistical confidence regions. 
The right panel shows that, for a multi-parameter fit,
the difference between the results with and
without the non-Gaussian errors is significantly suppressed due to the
parameter degeneracies. 
Note that our results in Fig.\ref{dist_w} is independent of the
assumed survey volume, because $[w/(w)_{\rm fid}-1] \propto V^{-1/2}$ from
 Eq.(\ref{deviation}) and $\sigma \propto V^{-1/2}$.
Although we show the result only for $w$, essentially the same
results are obtained for other parameters ($A_s,n_s,\Omega_b h^2,\Omega_c h^2$).

\begin{figure*}
\epsscale{0.8}
\plottwo{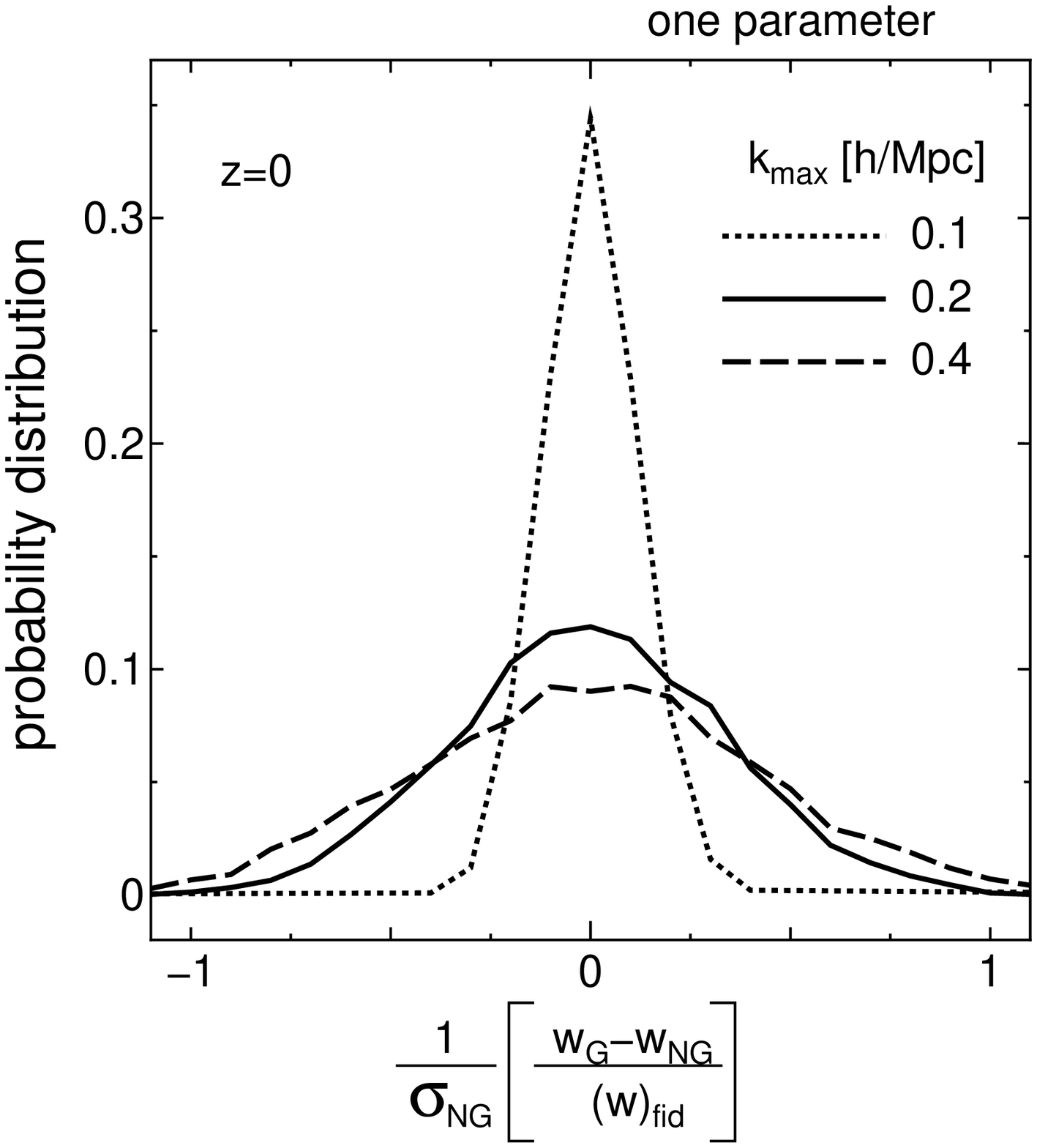}{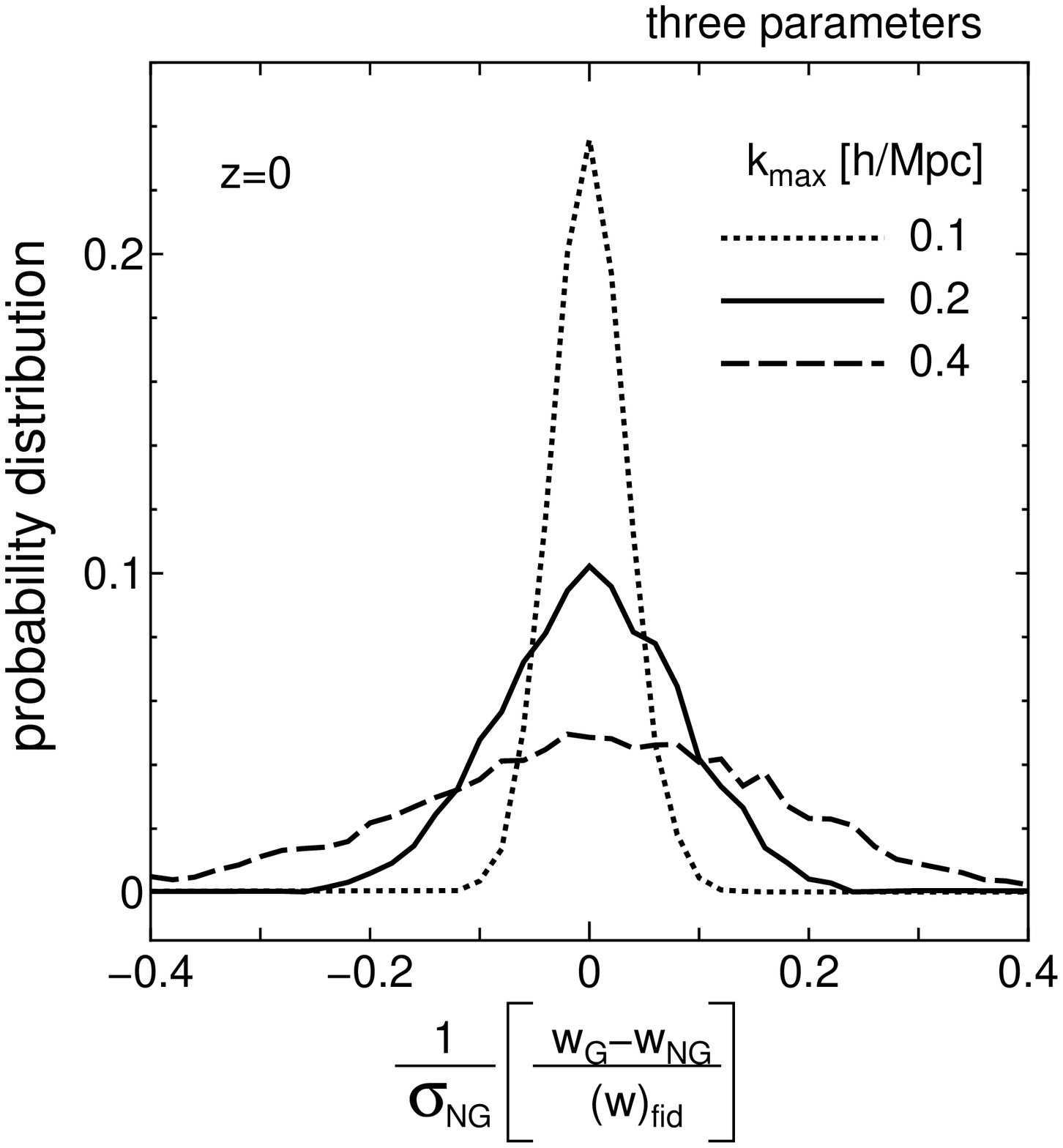}
\caption{As for the previous figures, but for
the distribution of the differences between the best-fit values of $w$ computed
 from Eq.~(\ref{deviation})
with and without non-Gaussian errors. 
The $\sigma_{\rm NG}$ is 
the $1 \sigma$ error computed from the Fisher matrix with the
 non-Gaussian errors. 
}
\label{dist_w_d}
\vspace*{0.5cm}
\end{figure*}

In Fig.\ref{dist_w_d}, we quantify how the best-fit values of $w$ are
systematically different when including or ignoring the non-Gaussian
errors. 
The horizontal axis is the difference between the best-fit parameters, 
 ($w_{\rm G}-w_{\rm NG})/w_{\rm fid}$, divided by the 1$\sigma$
 statistical confidence error derived by including the non-Gaussian
 errors. 
The left panel shows that the differences are smaller than the $1\sigma$
confidence regions, and the right panels 
show even much smaller
differences for the three-parameter fitting. Therefore we again conclude
that the non-Gaussian errors do not cause any significant bias in the
best-fit value compared to the statistical
confidence regions including the non-Gaussian errors. 

\section{Acoustic Peak Positions}

\begin{figure}
\vspace*{0.5cm}
\epsscale{0.9}
\plotone{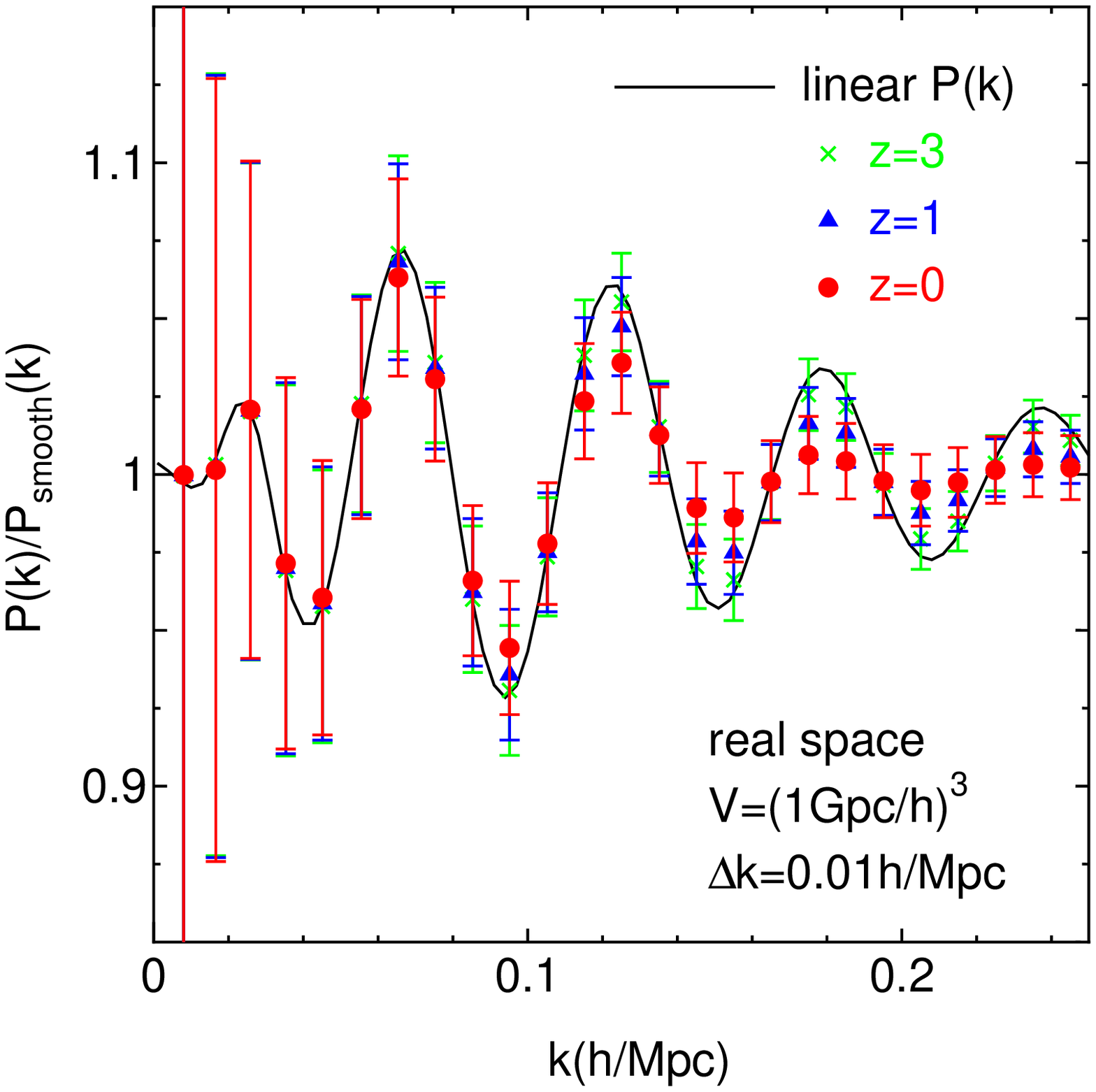}
\caption{The real-space 
power spectrum divided by the smoothed spectrum for $z=0$, $1$ and 
$3$, respectively. 
The symbols denote the mean of the power spectra among 5000
 realizations of ($1h/{\rm Gpc}$)$^3$ volume,
 while the error bars denote 
the $1 \sigma$ scatters. 
The solid curve is for the linear $P(k)$.
The BAO features are more erased at higher $k$ and at lower redshifts
 due to stronger nonlinearities.
}
\label{pk_per_ps}
\vspace*{0.5cm}
\end{figure}

\begin{figure}
\vspace*{0.5cm}
\epsscale{0.9}
\plotone{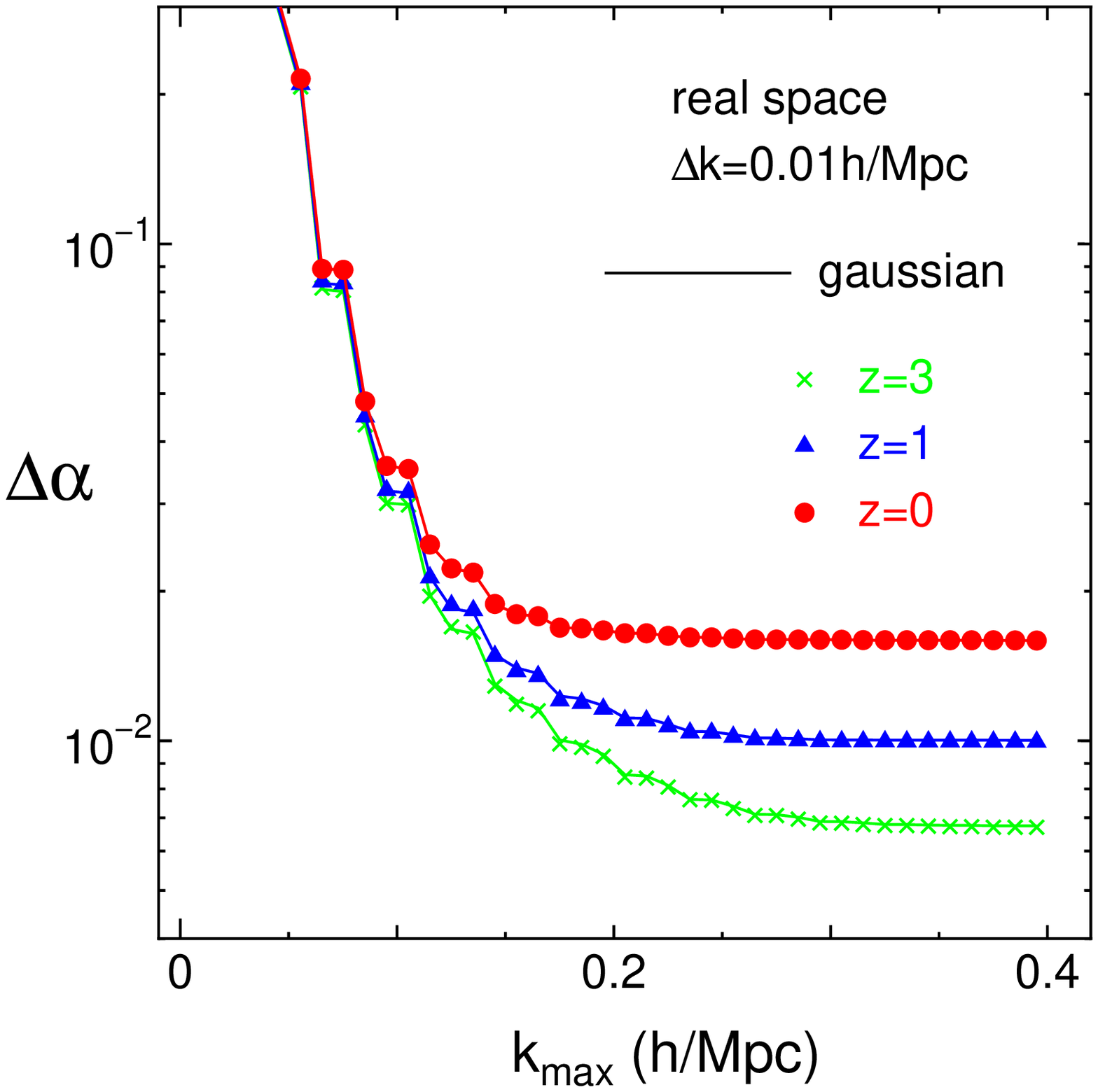}
\caption{The $1\sigma$ error on the BAO peak position parametrized by
 the stretch parameter $\alpha$ (see text for the definition).
The symbols are the results computed from simulations, while the solid
 curves, which are almost top on the symbols, show the errors computed
assuming the Gaussian errors. 
}
\label{fig_fisher_alpha}
\vspace*{0.5cm}
\end{figure}

A more robust method to constrain dark energy is using the BAO peak
positions. The BAO peak
positions are characterized basically by {\em one} parameter, the
stretch parameter (see below), and therefore the non-Gaussian errors may
significantly affect the accuracy of the peak position determination. Here we
use the distribution of the BAO peak positions obtained from our  
$5000$ realizations.
We employ the method developed in 
Percival et al. (2007) to estimate the peak positions (see also 
 Nishimichi et al. 2009). We first divide the measured $P(k)$ by a
 smooth model, which is constructed by 
adopting the cubic B-spline function to fit the binned power spectrum
over a range of wavenumbers binned with the width
$\Delta k=0.01h/$Mpc.

Fig.\ref{pk_per_ps} shows the  power spectrum divided by 
the smooth model: the data points are the average spectrum of 5000
realizations and the errors the $1\sigma$ variation ranges, 
$\Delta P(k) = {\rm cov}^{1/2}(k,k)$.
Clearly, the BAO features are smoothed out at larger $k$ and 
at lower redshifts due to stronger nonlinearities. 

To make parameter forecasts, 
let us define the ratio power spectrum, $R(k)$, as  
\beq
  R(k) = \frac{P(k)}{P_{\rm smooth}(k)}. 
\eeq
Then we can introduce  the stretch parameter $\alpha$ which 
characterizes a shift of the BAO peak phases via the transform 
$k \rightarrow \alpha k$
 in $R(k)$.
The power spectrum with the stretch parameter $\alpha$ is 
given as 
\beq
  P(k;\alpha)=P_{\rm smooth}(k) R(\alpha k),
\eeq
and $\alpha=1$ is the fiducial model.
The Fisher information matrix for the stretch parameter $\alpha$ is
  computed 
as 
\beq
 F_{\alpha \alpha} =  \sum_{k_{1,2}<k_{\rm max}}
 {\rm cov}^{-1} (k_1,k_2) \frac{dP(k_1;\alpha)}{d\alpha}
 \frac{dP(k_2;\alpha)}{d\alpha}.
\eeq
Since we focus on the BAO peak locations, we treat 
only 
$\alpha$ 
as a free parameter,
 and hence the precision of determining $\alpha$ 
for the given power spectrum measurement 
is given as
$\Delta \alpha =
 F^{-1/2}_{\alpha \alpha}$.

Fig.\ref{fig_fisher_alpha} shows the $1\sigma $
error, $\Delta \alpha$,
 as a function of $k_{\rm max}$ up to which the power spectrum 
information is included. 
The symbols are the results including the non-Gaussian errors in the
Fisher analysis, 
while the solid
 curves are for the Gaussian errors. 
The Gaussian error assumption appears to be 
valid
 even for large $k_{\rm max}$.
This is because the BAO features are erased at the weakly nonlinear
scales $k> 0.2h/$Mpc, where the non-Gaussian errors are more
significant.
There is little information on 
the acoustic scale at the nonlinear scales.
At $z=0$ the accuracy improves significantly around the first peak
 ($k \sim 0.06h/$Mpc) and the second peak ($k \sim 0.12h/$Mpc).
Hence almost all the information on the acoustic peaks are obtained for
 $k \lesssim 0.15h/$Mpc at $z=0$.
From Fig.\ref{fig_fisher_alpha}, $\Delta \alpha \sim 1\%$ can be achievable
 for a survey with 
$({\rm Gpc}/h)^3$ volume coverage.

\begin{figure}
\vspace*{0.5cm}
\epsscale{0.9}
\plotone{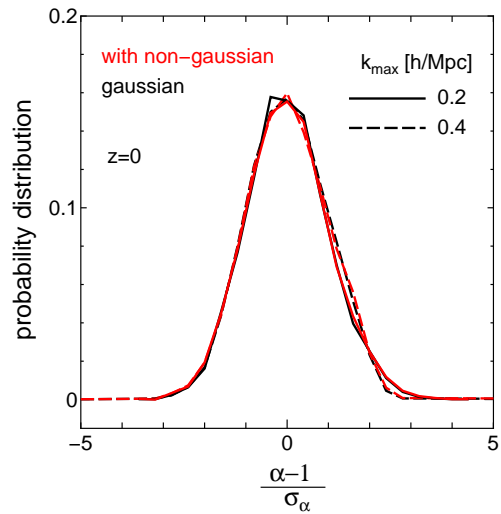}
\caption{The distribution of 
the shift parameter $\alpha$ for $k_{\rm max}=0.2$
 (solid curves) and $0.4h/$Mpc (dashed curves), respectively.
The horizontal axis $\alpha-1$ is divided by the $1 \sigma$ error 
computed from the Fisher matrix.
The red and black curves are the results with and without the
 non-Gaussian errors, respectively.
The distribution of $\alpha$ is well described by a 
 Gaussian distribution with the width of 
the 1$\sigma$ Fisher error. 
}
\label{fig_dist_alpha}
\vspace*{0.5cm}
\end{figure}

\begin{figure}
\vspace*{0.5cm}
\epsscale{0.9}
\plotone{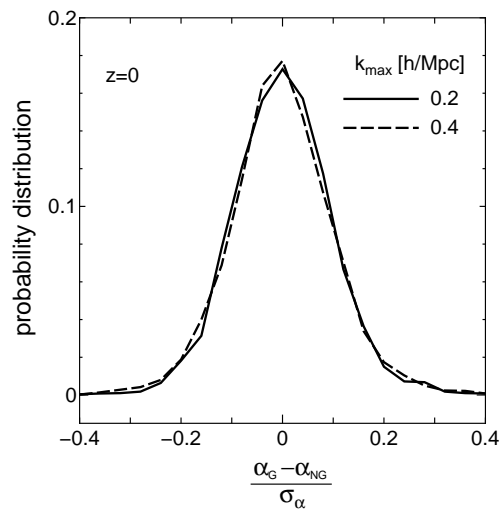}
\caption{The distribution of the differences between the best-fit values
 of $\alpha$ 
with or without including
the non-Gaussian covariance.
The solid (dashed) curve is for $k_{\rm max}=0.2 (0.4) h/$Mpc.
}
\label{fig_dist_alpha_d}
\vspace*{0.5cm}
\end{figure}

Finally, we investigate the distribution of best-fit $\alpha$
 among $5000$ realizations. Given the power spectrum measurement for 
the $i$-th realization, 
the $\chi^2$ for estimating $\alpha$ 
 is given as
\beqa
 \chi^2_i(\alpha) = \sum_{k_{1,2}<k_{\rm max}} {\rm cov}^{-1}(k_1,k_2)
 P_{\rm smooth}(k_1) P_{\rm smooth}(k_2)    \nonumber  \\ 
 \times \left[ R(\alpha k_1) - \hat{R}_i(k_1) \right]
 \left[ R(\alpha k_2) - \hat{R}_i(k_2) \right].  \hspace{1cm}
\label{chi2_alpha}
\eeqa
Here, 
 we have used the power spectrum $P_i(k)$ for the i-th realization,
 not the mean $P(k)$,
 to obtain the smooth power spectrum
 in the denominator of the ratio $\hat{R}_i(k)$,

Fig.\ref{fig_dist_alpha} shows the distribution of 
the best-fit shift parameters
$\alpha$ for the cases of 
 $k_{\rm max}=0.2 $ and $0.4h/$Mpc.
The horizontal axis is $\alpha-1$ divided by the $1\sigma$ error
 $\sigma_\alpha$
 obtained by the Fisher matrix.
The red curves are the results 
including the non-Gaussian errors, while the black
 curves are for the Gaussian errors.
The two results are indeed very similar. 
The distribution is well described by a Gaussian function with the width
given by the $1\sigma$ Fisher error. 
A recent work by Seo et al. (2009) compares the distribution of the acoustic scales 
with the result from
the Fisher matrix analysis.
Although their analysis is slightly different from ours (they use a
 fitting formula for the non-linear $P(k)$ 
when calculating the acoustic scale), their results are 
broadly consistent with the results shown here.

Fig.\ref{fig_dist_alpha_d} shows the difference between the best-fit 
shift parameters when including or ignoring the non-Gaussian errors
 in Eq.(\ref{chi2_alpha}).
The difference is smaller than $\sim 0.2 \sigma$.
Hence we conclude that the non-Gaussian covariance 
does not cause a substantial systematic error in the BAO peak determination. 

\section{Discussion and Conclusion}

We have studied the effects of the non-Gaussian error on the parameter
 estimations and the distribution of the best-fit parameters
 for the cosmological parameters (in section 3)
 and the acoustic scale position (in section 4).
We have found that the non-Gaussian error is important for the parameter
 errors if there is only one fitting parameter.
The measurement error degrades up to factor 5 for $k_{\rm max}=0.4$. 
However, if there are more than two parameters, the impact of the 
non-Gaussian errors are insignificant due to severe parameter
 degeneracies in the matter power spectrum. 
For the acoustic scale, the non-Gaussian errors do not affect the
 acoustic scale determination, even though  
there is only one fitting parameter.
This is because that the acoustic scale is determined mainly by the
linear scales 
where the non-Gaussian covariance is not important.

Throughout this paper, we discussed the covariance matrix of the matter
 power spectrum. However, for real galaxy survey, we should include the
 effects of the halo and galaxy bias.
However addressing this issue is not easy at present, because we need
 the large number of realizations of high resolution simulations
 including halo and galaxy formation.
For large scale ($k \lesssim 0.1h/$Mpc), the covariance of the halo power
 spectrum is consistent with the Gaussian error with the shot noise
 term (Angulo et al. 2008; Smith 2009). 
Because, in the linear regime, only the sample variance dominates
 the covariance.
For small scale ($k > 0.1h/$Mpc), Smith (2009) recently showed that the
 covariance is larger than the Gaussian error prediction and the higher
 mass halo have stronger covariance due to the non-linear gravitational
 mode coupling.
He compared the cluster-sized halos in the two mass ranges
 ($M>10^{14} M_\odot$ and $1 \times
 10^{13} M_\odot < M < 2 \times 10^{13} M_\odot$), and hence we expect
 the galactic halo ($M \lesssim 10^{13} M_\odot$) would have a weaker
 covariance.

In our previous paper (T09), we compare the power spectrum covariance
 in numerical simulations with an analytical models such as
 perturbation theory and halo model.
We calculated the diagonal and off-diagonal terms of the covariance matrix
 and the signal-to-noise ratio in the both models, 
 and found that the halo model reasonably well reproduces the simulation
 results.
Several authors also compare them and reached the same conclusion
 (e.g. Cooray \& Hu 2001; Neyrinck et al. 2006; Neyrinck \& Szapudi 2007;
 Sato et al. 2009).

Our simulation results of the $5000$ power spectra $P_{\rm i}(k)$, the
 derivative of $P(k)$ with respect to the cosmological parameters,
 $\partial P(k)/\partial x_{\rm j}$, and the covariance matrix 
 ${\rm cov}(k_1,k_2)$ are available as numeric tables
 upon request (contact takahasi@cc.hirosaki-u.ac.jp).

\appendix
\section{Variance of the Covariance Matrix for the Gaussian Density 
 Fluctuations}

The covariance matrix estimated from $N_{\rm r}$ realizations is given by,
\beqa
  {\rm cov}(k_1,k_2;N_{\rm r}) = && \frac{1}{N_{\rm r}} \sum_i
 \left[ \hat{P}_i(k_1) - \bar{P}(k_1) \right]  \left[ \hat{P}_i(k_2)
 - \bar{P}(k_2) \right], 
  \nonumber \\
 = && \frac{1}{N_{\rm r}} \sum_i \hat{P}_i(k_1)  \hat{P}_i(k_2)
  - \frac{1}{N_{\rm r}} \sum_i \hat{P}_i(k_1)
  \frac{1}{N_{\rm r}} \sum_j \hat{P}_j(k_2).  
\eeqa
The variance of the covariance is given by,
\beqa
 {\rm var} \left[ {\rm cov}(k_1,k_2;N_{\rm r}) \right] \equiv && \langle
 {\rm cov}^2(k_1,k_2;N_{\rm r}) \rangle -
 \langle {\rm cov}(k_1,k_2;N_{\rm r}) \rangle^2 \\ \nonumber
 = && \frac{1}{N_{\rm r}^2} \sum_{i,j} \left[ \langle \hat{P}_i(k_1)
 \hat{P}_i(k_2) \hat{P}_j(k_1) \hat{P}_j(k_2) \rangle - \langle \hat{P}_i(k_1)
 \hat{P}_i(k_2) \rangle \langle \hat{P}_j(k_1) \hat{P}_j(k_2) \rangle \right]
 \\ \nonumber
 - && \frac{2}{N_{\rm r}^3} \sum_{i,j,k} \left[ \langle \hat{P}_i(k_1)
 \hat{P}_i(k_2) \hat{P}_j(k_1) \hat{P}_k(k_2) \rangle - \langle \hat{P}_i(k_1)
 \hat{P}_i(k_2) \rangle \langle \hat{P}_j(k_1) \hat{P}_k(k_2) \rangle \right]
 \\ \nonumber
 + && \frac{1}{N_{\rm r}^4} \sum_{i,j,k,l} \left[ \langle \hat{P}_i(k_1)
 \hat{P}_j(k_2) \hat{P}_k(k_1) \hat{P}_l(k_2) \rangle - \langle \hat{P}_i(k_1)
 \hat{P}_j(k_2) \rangle \langle \hat{P}_k(k_1) \hat{P}_l(k_2) \rangle \right],
\eeqa
The above equation further reduces to
\beqa
  {\rm var} \left[ {\rm cov}(k_1,k_2;N_{\rm r}) \right] = &&
  \frac{1}{N_{\rm r}} \left[ \langle \hat{P}^2(k_1) \hat{P}^2(k_2) \rangle
 - \langle \hat{P}(k_1) \hat{P}(k_2) \rangle^2  \right]  \\ \nonumber
 -&& \frac{2}{N_{\rm r}} \left[ \langle \hat{P}^2(k_1) \hat{P}(k_2) \rangle
 \langle \hat{P}(k_2) \rangle -  \langle \hat{P}(k_1) \hat{P}(k_2) \rangle
  \langle \hat{P}(k_1) \rangle \langle \hat{P}(k_2) \rangle
 + (k_1 \leftrightarrow k_2) \right]  \\ \nonumber
 +&& \frac{1}{N_{\rm r}} \left[ \left( \langle \hat{P}^2(k_1) \rangle
 - \langle \hat{P}(k_1) \rangle^2 \right) \langle \hat{P}(k_2) \rangle^2 
 + (k_1 \leftrightarrow k_2) \right]  \\ \nonumber
 +&& \frac{2}{N_{\rm r}} \left[ \langle \hat{P}(k_1) \hat{P}(k_2) \rangle
 - \langle \hat{P}(k_1) \rangle  \langle \hat{P}(k_2) \rangle \right]
 \langle \hat{P}(k_1) \rangle  \langle \hat{P}(k_2) \rangle  
\eeqa
Here we ignored the terms of the order of $(1/N_{\rm r})^2$ and higher.

For the Gaussian density fluctuations, the $n$-th moments $\langle \hat{P}^n
 \rangle$ can be obtained using the probability distribution function of
 $P(k)$ (the chi-squared distribution function, see Eq.(B1) in T09).
For the diagonal parts, we have
\beq
  {\rm var} \left[ {\rm cov}(k_1,k_1;N_{\rm r}) \right] =
  \frac{1}{N_{\rm r}} \left( \frac{8}{N_k^2} + \frac{48}{N_k^3} \right)
  \simeq \frac{8}{N_{\rm r} N_k^2} P^4(k)
\eeq
Hence we have
\beq
  \frac{{\rm var} \left[ {\rm cov}(k_1,k_1;N_{\rm r}) \right]}
 {{\rm cov}^2(k_1,k_1)} = \frac{2}{N_{\rm r}}
\eeq
which is consistent with our numerical finding in our previous paper
 (see left panel of Fig.12 in T09).
Hence if we need $10\% (5 \%)$ accuracy in the covariance, we have to
 prepare $200 (800)$ realizations.

\begin{figure}
\epsscale{0.45}
\plotone{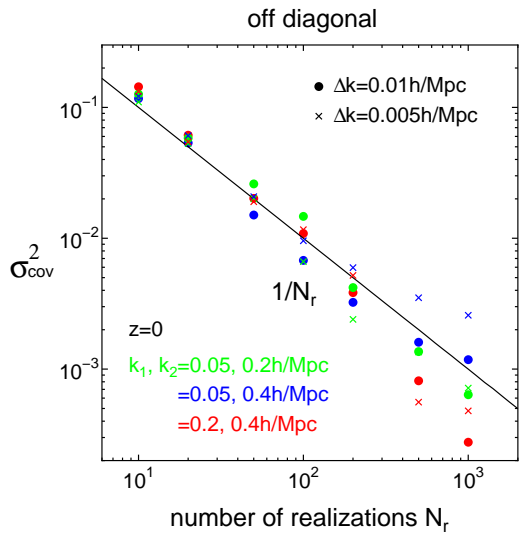} 
\caption{ The dispersions among the power spectrum
covariances each of which is estimated from the $N_{\rm r}$ realizations
(a subset of the while 5000 realizations), as a function of $N_{\rm r}$.
In the vertical axis, the dispersion is defined in Eq.~(\ref{var_cov_def}).
The panel shows the result for the off-diagonal
 parts for varying the wavenumber bins and the bin widths.  The
color symbols are the simulation results, while the solid curves denote the
 theoretical prediction for the Gaussian density fluctuation.
The plots explicitly
show that the power spectrum covariances are estimated at a sub-percent
level accuracy by using our whole 5000 realizations.  }
\label{fig_convergence_cov} \hspace{0.5cm}
\end{figure}

For the off-diagonal parts we have
\beq
 {\rm var} \left[ {\rm cov}(k_1,k_2;N_{\rm r}) \right] =
 \frac{1}{N_{\rm r}} \frac{2}{N_{k_1}} \frac{2}{N_{k_2}}
 P^2(k_1) P^2(k_2)
\label{var_cov_offdiag}
\eeq  
Let us define the relative errors as,
\beq
  \sigma_{\rm cov}^2 = \frac{{\rm var} \left[{\rm cov}(k_1,k_2;N_{\rm r})
 \right]} {{\rm cov}(k_1,k_1)~{\rm cov}(k_2,k_2)}.
\label{var_cov_def}
\eeq
In Fig.\ref{fig_convergence_cov}, we show the relative errors
 $\sigma_{\rm cov}^2$ in Eq.(\ref{var_cov_def}) as a function of
 $N_{\rm r}$.
The solid line is the theoretical prediction in Eq.(\ref{var_cov_def}), and 
 the symbols are the results of our numerical simulation for various scales
 ($k=0.05,0.2,0.4h/$Mpc) and bin-width ($\Delta k=0.005,0.01h/$Mpc).
Analytical results in Eq.(\ref{var_cov_offdiag}) fit our simulation data well.

\acknowledgments
We would like to thank Daniel Eisenstein and Hee-Jong Seo for useful
 comments and discussions. 
This work is supported in part by Grant-in-Aid for Scientific Research
on Priority Areas No. 467 ``Probing the Dark Energy through an
Extremely Wide and Deep Survey with Subaru Telescope'',
 by the Grand-in-Aid for the Global COE Program
 ``Quest for Fundamental Principles in the Universe: from Particles
 to the Solar System and the Cosmos'' from the Ministry of Education,
 Culture, Sports, Science and Technology (MEXT) of Japan, 
by the World Premier
International Research Center Initiative of MEXT of Japan,
 by the Mitsubishi Foundation,
and by Japan Society for Promotion of Science (JSPS) Core-to-Core
Program ``International Research Network for Dark Energy'', and by
Grant-in-Aids for Scientific Research
(Nos.~18740132,~18540277,~18654047).  I.K., T.N. and S.S. are
supported by Grants-in-Aid for Japan Society for the Promotion of
Science Fellows.
A.T. is supported in part by a Grants-in-Aid for Scientific
 Research from the JSPS (No. 21740168).

\clearpage

\end{document}